\documentclass[10pt,conference]{IEEEtran}

\usepackage{graphicx}

\usepackage{dashrule}

\usepackage[linesnumbered,ruled,vlined]{algorithm2e}
\usepackage{algpseudocode}
\usepackage{amsmath,amssymb}
\usepackage{multirow}
\usepackage{xspace}
\usepackage{hhline}

\usepackage{url}
\DeclareFixedFont{\ttb}{T1}{txtt}{bx}{n}{10} 
\DeclareFixedFont{\ttm}{T1}{txtt}{m}{n}{10}  

\usepackage{color}
\definecolor{deepblue}{rgb}{0,0,0.5}
\definecolor{deepred}{rgb}{0.6,0,0}
\definecolor{deepgreen}{rgb}{0,0.5,0}

\usepackage{listings}

\newcommand\pythonstyle{\lstset{
		language=Python,
		basicstyle=\ttm,
		otherkeywords={self},             
		keywordstyle=\ttb\color{deepblue},
		emph={MyClass,__init__},          
		emphstyle=\ttb\color{deepred},    
		stringstyle=\color{deepgreen},
		frame=tb,                         
		showstringspaces=false,            %
		numbers=left
}}

\makeatletter
\def\lst@makecaption{%
  \def\@captype{table}%
  \@makecaption
}
\makeatother

\lstnewenvironment{python}[1][]
{
	\pythonstyle
	\lstset{#1}
}
{}

\usepackage[text size=small, backgroundcolor=blue!30,bordercolor=blue,linecolor=blue]{todonotes}

\newcommand{\substepseparator}{\hspace{1cm}}

\newcommand{\swat}{SWaT\xspace}
\newcommand{\wadi}{WADI\xspace}
\newcommand{\cps}{CPS\xspace}

\begin{document}

\title{Learning-Guided Network Fuzzing for\\ Testing Cyber-Physical System Defences}

\author{\IEEEauthorblockN{Yuqi Chen\IEEEauthorrefmark{1}, Christopher M. Poskitt\IEEEauthorrefmark{1}, Jun Sun\IEEEauthorrefmark{2}, Sridhar Adepu\IEEEauthorrefmark{1}, and Fan Zhang\IEEEauthorrefmark{3}}
\\
\IEEEauthorblockA{\IEEEauthorrefmark{1}Singapore University of Technology and Design, Singapore}
\IEEEauthorblockA{\IEEEauthorrefmark{2}Singapore Management University, Singapore}
\IEEEauthorblockA{\IEEEauthorrefmark{3}Zhejiang University, Zhejiang Lab, and Alibaba-Zhejiang University Joint Institute of Frontier Technologies, China}}

\maketitle

\begin{abstract}
	The threat of attack faced by cyber-physical systems (CPSs), especially when they play a critical role in automating public infrastructure, has motivated research into a wide variety of attack defence mechanisms. Assessing their effectiveness is challenging, however, as realistic sets of attacks to test them against are not always available. In this paper, we propose \emph{smart fuzzing}, an automated, machine learning guided technique for systematically finding `test suites' of CPS network attacks, without requiring any knowledge of the system's control programs or physical processes. Our approach uses predictive machine learning models and metaheuristic search algorithms to guide the fuzzing of actuators so as to drive the CPS into different unsafe physical states. We demonstrate the efficacy of smart fuzzing by implementing it for two real-world CPS testbeds---a water purification plant and a water distribution system---finding attacks that drive them into 27 different unsafe states involving water flow, pressure, and tank levels, including six that were not covered by an established attack benchmark. Finally, we use our approach to test the effectiveness of an invariant-based defence system for the water treatment plant, finding two attacks that were not detected by its physical invariant checks, highlighting a potential weakness that could be exploited in certain conditions.
\end{abstract}


\section{Introduction}\label{sec:introduction}

Cyber-physical systems~(CPSs) are characterised by computational elements and physical processes that are deeply intertwined, each potentially involving different spatial and temporal scales, modalities, and interactions~\cite{US-NSF18a}. These complex systems are now ubiquitous in modern life, with examples found in fields as diverse as aerospace, autonomous vehicles, and medical monitoring. CPSs are also commonly used to automate aspects of critical civil infrastructure, such as water treatment or the management of electricity demand~\cite{Rajkumar-et_al10a}. Given the potential to cause massive disruption, such systems have become prime targets for cyber attackers, with a number of successful cases reported in recent years~\cite{Leyden16a,ICS-Cert-Alert16a}.

This pervasive threat faced by CPSs has motivated research and development into a wide variety of attack defence mechanisms, including techniques based on anomaly detection~\cite{Cheng-Tian-Yao17a,Harada-et_al17a,Inoue-et_al17a,Pasqualetti-Dorfler-Bullo11a,Aggarwal-et_al18a,Aoudi-et_al18a,He-et_al18a,Kravchik-Shabtai18a,Lin-et_al18a,Narayanan-Bobba18a,Schneider-Boettinger18a}, fingerprinting~\cite{Ahmed-et_al18a,Ahmed-et_al18b,Gu-et_al18a,Kneib-Huth18a}, and monitoring conditions or physical invariants~\cite{Cardenas-et_al11a,Adepu-Mathur16a,Adepu-Mathur16b,Chen-Poskitt-Sun16a,Adepu-Mathur18b,Chen-Poskitt-Sun18a,Choi-et_al18a,Giraldo-et_al18a}. The practical utility of these different countermeasures ultimately depends on how effective they are at their principal goal: \emph{detecting and/or preventing attacks}. Unfortunately, assessing this experimentally is not always straightforward, even with access to an actual CPS, because of the need for realistic attacks to evaluate them against. Herein lies the problem: \emph{where exactly} should such a set of attacks come from?

One possible solution is to use existing attack benchmarks and datasets, as have been made available by researchers for different CPS testbeds~\cite{CPS-Datasets,Goh-et_al16a}, and as have been used in the evaluation of different countermeasures, e.g.~\cite{Inoue-et_al17a,Kravchik-Shabtai18a,Lin-et_al18a}. Across these examples, attackers are typically assumed to have compromised the communication links to some extent, and thus can manipulate the sensor readings and actuator commands exchanged across the network. The attacks within these benchmarks, however, are manually constructed, whether by engineers with sufficient expertise in the CPS, or through invited hackathons to discover new attacks from those without insider bias~\cite{Adepu-Mathur18a}. Both methods are inherently time consuming, and difficult to generalise: constructing a similar benchmark for a new CPS essentially requires starting from scratch.

In this paper, we propose \emph{smart fuzzing}, an automated, machine learning~(ML) guided approach for constructing `test suites' (or benchmarks) of CPS network attacks, without requiring any specific system expertise other than knowing the normal operational ranges of sensors. Our technique uses predictive machine learning models and metaheuristic search to intelligently fuzz actuator commands, and systematically drive the system into different categories of unsafe physical states. Smart fuzzing consists of two broad steps. First, we \emph{learn} a model of the CPS by training ML algorithms on physical data logs that characterise its normal behaviour. The learnt model can be used to predict how the current physical state will evolve with respect to different actuator configurations. Second, we \emph{fuzz} the actuators over the network to find attack sequences that drive the system into a targeted unsafe state. This fuzzing is guided by the learnt model: potential manipulations of the actuators are searched for (e.g.~with a genetic algorithm~\cite{Goldberg89a}), and then the model predicts which of them would drive the CPS closest to the unsafe state.

Our design for smart fuzzing was driven by four key requirements. First, that it should be \emph{general}, in the sense that it can be implemented for different CPSs and a variety of sensors and actuators. Second, that the approach should be \emph{comprehensive}, in that the suites of attacks it constructs should systematically cover different categories of sensed physical properties, rather than just a select few. Third, that it should be \emph{efficient}, with each attack achieving its goal quickly, posing additional challenge for countermeasures. Finally, that it should be \emph{practically useful}, in that it is straightforward to implement for real CPSs without any formal specification or specific technical expertise, and that the `test suites' of attacks are of comparable quality to expert-crafted benchmarks, thus a reasonable basis for assessing attack defence mechanisms.

To evaluate our approach against these requirements, we implemented it for two CPS testbeds. First, the Secure Water Treatment~(\swat) testbed~\cite{SWaT-Reference}, a fully operational water treatment plant consisting of 42 sensors and actuators, able to produce five gallons of drinking water per minute. Second, the Water Distribution~(\wadi) testbed~\cite{Ahmed-et_al17a}, a scaled-down version of a typical water distribution network for a city, built for investigating attacks on consumer water supplies with respect to patterns of peak and off-peak demand. The designs of these testbeds were based on real-world industrial purification plants and distribution networks, and thus reflect many of their complexities. We found that smart fuzzing could automatically identify suites of attacks that drove these CPSs into 27 different unsafe states involving water flow, pressure, tank levels, and consumer supply. Furthermore, it covered six unsafe states beyond those in an established expert-crafted benchmark~\cite{Goh-et_al16a}. Finally, we evaluated the utility of smart fuzzing for testing attack defence mechanisms by launching it with \swat's invariant-based monitoring system enabled~\cite{Adepu-Mathur18b,Adepu-Mathur16c}. Our approach was able to identify two attacks that evaded detection by its physical invariant checks, highlighting a potential weakness that could be exploited by attackers with the capabilities to bypass its other conditions.

\substepseparator

\noindent\textbf{Summary of Contributions.} We propose smart fuzzing, which to the best of our knowledge, is the first general black-box technique for automatically constructing test suites (or benchmarks) of network attacks for different CPSs. While fuzzing itself is not a new idea, our work differs in its focus on CPS actuators, and the use of ML models to guide the process and find attacks covering multiple different goals. We implemented smart fuzzing for two real-world CPS testbeds, identifying attacks that drive them into 27 different unsafe states, including six that were not present in an expert-crafted benchmark. Using the approach, we discovered two potential exploits in an established invariant-based monitoring system, suggesting the potential utility of smart fuzzing for testing CPS attack defence mechanisms.

For researchers, our work provides a general way of constructing attack benchmarks for CPSs, which in turn could be used in the experimental evaluation of novel attack defence mechanisms. For plant engineers, it provides a practical means of assessing a system's robustness against a variety of network attacks spanning several different goals. For the ML and search communities, it demonstrates that existing techniques can be used together to overcome the complexities of real CPSs.

\section{Background and Motivational Example}\label{sec:swat_testbed} 

Here, we clarify our assumptions of CPSs and fuzzing, before introducing our real-world CPS case studies, \swat and \wadi. We discuss an example of smart fuzzing on \swat.

\substepseparator

\noindent\textbf{CPSs and Fuzzing.} We define CPSs as systems in which algorithmic control and physical processes are tightly integrated. Concretely, we assume that they consist of computational elements (the `cyber' part) such as programmable logic controllers~(PLCs), distributed over a network, and interacting with their processes via sensors and actuators (the `physical' part). The operation of a CPS is controlled by its PLCs, which receive readings from sensors that observe the physical state, and then compute appropriate commands to send along the network to the relevant actuators. We assume that the sensors read continuous data (e.g.~temperature, pressure, flow) and that the states of the actuators are discrete.

These characteristics together make CPSs very difficult to reason about: while individual control components (e.g.~PLC programs) may be simple in isolation, reasoning about the behaviour of the whole system can only be done with consideration of how its physical processes evolve and interact. This often requires considerable domain-specific expertise beyond the knowledge of a typical computer scientist, which is one of our principal motivations for achieving full automation.

Fuzzing, which plays a key role in our solution, is in general an automated testing technique that attempts to identify potential crashes or assertion violations by generating diverse and unexpected inputs for a given system~\cite{Takanen-Demott-Miller18a}. Many of the most well-known tools perform fuzzing on programs, e.g.~\cite{Cha-Woo-Brumley15a,Zalewski}, but in the context of CPSs, we consider fuzzing at the \emph{network} level. In particular, we consider the fuzzing of their actuators with commands that did not originate from the PLCs (and in fact override any valid commands that they are trying to send). Furthermore, the goal of our fuzzing differs in that we are trying to drive physical sensors out of their safe ranges, using an underlying method that is ML-guided.

\substepseparator

\begin{figure}[!t]
	\centering
	\includegraphics[width=0.8\linewidth]{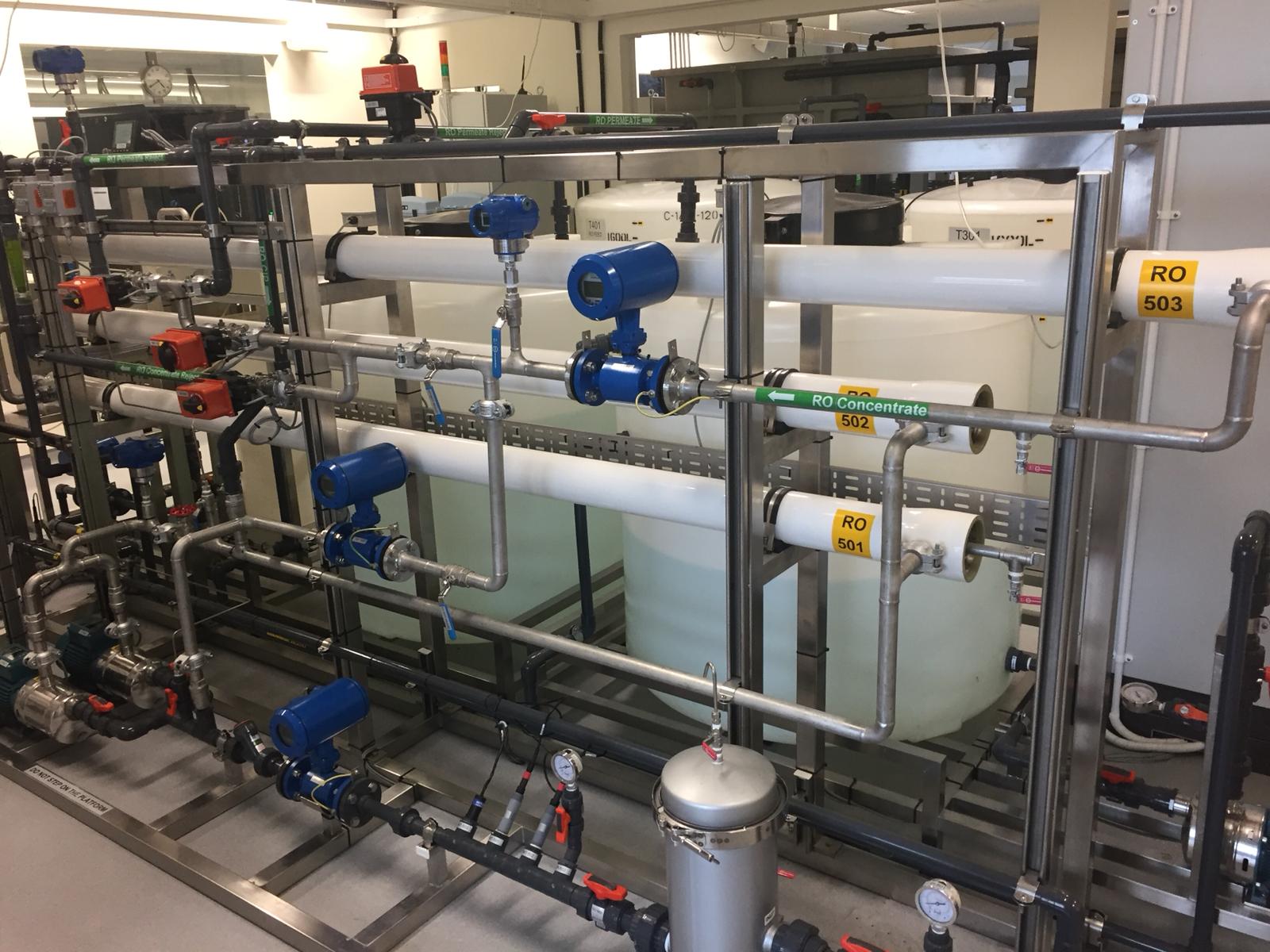}
	\caption{The Secure Water Treatment (\swat) testbed}
	\label{fig:swat_testbed}
\end{figure}

\noindent\textbf{\swat and \wadi Testbeds.} Two CPSs that satisfy the aforementioned assumptions, and thus will form the case studies of this paper, are Secure Water Treatment (\swat)~\cite{SWaT-Reference} and Water Distribution (\wadi)~\cite{Ahmed-et_al17a}, testbeds built for cyber-security research. \swat (Figure~\ref{fig:swat_testbed}) is a fully operational water treatment plant with the capability of producing five gallons of safe drinking water per minute, whereas \wadi is a water distribution network that supplies consumers with 10 gallons of drinking water per minute. The testbeds are scaled-down versions of actual water treatment and distribution plants in a city, and exhibit many of their complexities.

\swat treats water across six distinct but co-operating stages, involving a variety of complex chemical processes, such as de-chlorination, reverse osmosis, and ultrafiltration. Each stage is controlled by a dedicated Allen-Bradley ControlLogix PLC, which communicates with the sensors and actuators relevant to that stage over a ring network, and with other PLCs over a star network, using an EtherNet/IP protocol specific to the manufacturer. Each PLC cycles through its program, computing the appropriate commands to send to actuators based on the latest sensor readings received as input. The system consists of 42 sensors and actuators in total, with sensors monitoring physical properties such as tank levels, flow, pressure, and pH values, and actuators including motorised valves (for opening an inflow pipe) and pumps (for emptying a tank). A historian regularly records the sensor readings and actuator commands during \swat's operation, facilitating data logs for offline analyses and machine learning. SCADA software and tools developed by Rockwell Automation are available to support analysis and experimentation.

WADI consists of three distinct processes each controlled by a National Instruments PLC. The first stage consists of two 2500 litre water tanks, which receive water from external sources. Water from these tanks feed through to two elevated reservoirs in the second stage, which then supply six consumer tanks based on a pre-set pattern of demand (e.g.~peak and off-peak usage). After meeting each consumer's demand, water is drained to the return water tank in the third stage. This water can then be pumped back to stage one for re-use.

The sensors in both testbeds are associated with manufacturer-defined ranges of \emph{safe} values, which in normal operation, they are expected to remain within. If a sensor reports a (true) reading outside of this range, we say the physical state of the CPS has become \emph{unsafe}. If a level indicator transmitter, for example, reports that the tank in stage one has become more than a certain percentage full (or empty), then the physical state has become unsafe due to the risk of an overflow (resp.~underflow). Unsafe pressure states indicate the risk of a pipe bursting, and unsafe levels of water flow indicate the risk of possible cascading effects in other parts of the system. In \wadi, unsafe flow levels may also indicate that a particular consumer's water supply has been compromised.

A number of countermeasures have been developed to prevent \swat or \wadi from entering unsafe states. Ghaeini and Tippenhauer~\cite{Ghaeini-Tippenhauer16a}, for example, monitor the network traffic with a hierarchical intrusion detection system, and Ahmed et al.~\cite{Ahmed-et_al18a,Ahmed-et_al18b} detect attacks by fingerprinting sensor and process noise. Other approaches learn models from physical data logs, and use them to evaluate whether or not the current state represents normal behaviour or not: most (e.g.~\cite{Inoue-et_al17a,Kravchik-Shabtai18a,Goh_et-al17a}) use unsupervised learning to construct these models, although Chen et al.~\cite{Chen-Poskitt-Sun16a,Chen-Poskitt-Sun18a} use supervised learning by automatically seeding faults in the control programs (of a high-fidelity simulator). Adepu and Mathur~\cite{Adepu-Mathur16a,Adepu-Mathur16b,Adepu-Mathur18b} systematically and manually derive a comprehensive set of physics-based invariants and other conditions that relate the states of actuators and sensors, with any violations of them during operation reported. This attack defence mechanism has been successfully deployed in professional \swat hackathons, detecting 14 out of the 16 physical process attacks devised by experienced practitioners from academia and industry~\cite{Adepu-Mathur18a}. Feng et al.~\cite{Feng-et_al19a} also generate invariants, but use an approach based on learning and data mining that can capture noise in sensor measurements more easily than manual approaches.

\substepseparator

\begin{figure}[!t]
	\centering
	\includegraphics[width=0.8\linewidth]{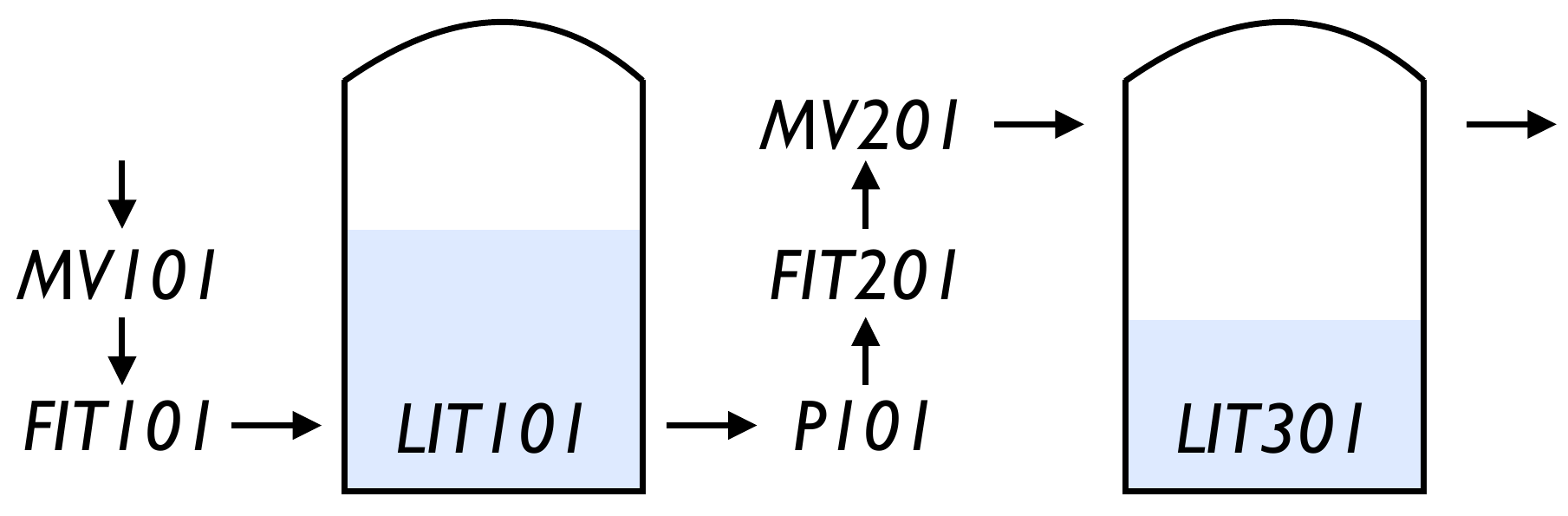}
	\caption{Some interconnected components of \swat's first three stages}
	\label{fig:lit101_example}
\end{figure}

\noindent\textbf{Smart Fuzzing Example.} To illustrate how our approach works in practice, we informally describe how it is able to automatically find an attack for overflowing a tank in the first stage of \swat. Figure~\ref{fig:lit101_example} depicts the relationship between some interconnected components across the first three stages. It includes some sensors: Level Indicator Transmitters~(LITs) for reporting the levels of different tanks; and Flow Indicator Transmitters~(FITs) for reporting the flow of water in some pipes. It also includes some actuators: Motorised Valves~(MVs), which if open, allow water to pass through; and a Pump~(P101), which if on, pumps water out of the preceding tank. The physical flow of water throughout this subpart of the system is controlled by some inter-communicating PLCs. If the value of LIT301 becomes too low, for example, the PLC controlling valve MV201 will open it. Furthermore, the PLC controlling pump P101 will switch it on to pump additional water through, causing the value of LIT301 to rise.

Before launching our smart fuzzer, two choices must be made: first, what is the \emph{attack goal} (characterised as a fitness function); and second, which search algorithm should be used to identify actuator configurations that achieve it? As our goal is to overflow the tank monitored by LIT101, we must define a fitness function on the sensor readings that is maximised as we get closer to overflowing the tank. A simple function achieving this takes as input a vector of predicted sensor states $\langle \mathrm{LIT101}, \mathrm{LIT301}, \dots, \mathrm{FIT101}, \mathrm{FIT201}, \dots \rangle$ and returns simply the value of LIT101. As our search strategy, we choose to randomly search over the space of actuator configurations. At this point, the user has nothing more to~set~up.

Upon launching the fuzzer, several different configurations of the system's actuators are (randomly) generated. For each set of configurations, a model is used to \emph{predict} the future sensor readings that they would lead to. Our fitness function is maximised by future states with the highest LIT101 readings, which in Figure~\ref{fig:lit101_example}, would result from a configuration where MV101 is open, P101 is off, P102 is off, and P601 is on (as water will be flowing into the tank, but the pumps will not be removing it). Note that P102 and P601 are not pictured: the former is a backup pump (redundancy for P101); the latter is a pump for driving in water from stage 6. With suitable actuator configurations identified, the relevant commands to apply them are spoofed over the network: $\mathtt{OPEN\ MV101}$, $\mathtt{OFF\ P101}$, $\mathtt{OFF\ P102}$, and $\mathtt{ON\ P601}$. Eventually, LIT101 will enter its upper unsafe range (risk of overflow), i.e.~the attack goal.

\section{How Smart Fuzzing Works} \label{sec:overview_of_approach}\label{sec:implementing_smart_fuzzing}

\begin{figure}[!t]
	\centering
	\includegraphics[width=0.8\linewidth]{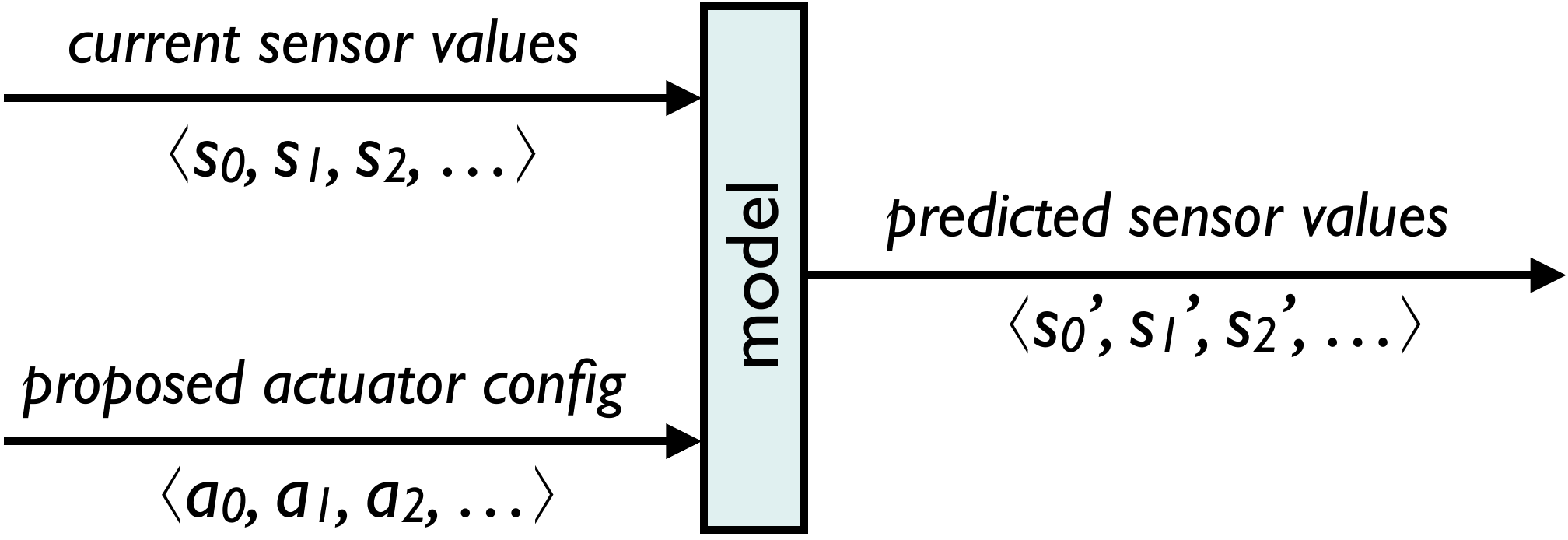}
	\caption{Inputs/outputs of a learnt model}
	\label{fig:model_input_output}
\end{figure}

Our approach for automatically finding network attacks on CPSs consists of two broad steps in turn: \emph{learning} and \emph{fuzzing}. In the first step, we learn a model of the \cps that can predict the effects of actuator configurations on the physical state, as summarised in Figure~\ref{fig:model_input_output}. The model takes as input the current readings of all sensors and a proposed configuration of the actuators, returning as output a prediction of the sensor readings that would result from adopting that configuration for a fixed time interval. The idea is that this model can later be used to analyse different potential actuator configurations, and help inform which of them is likely to drive the system closer to a targeted unsafe state. To learn this model, we extract a time series of sensor and actuator data from the system logs and train a suitable machine learning algorithm.

\begin{figure}[!t]
	\centering
	\includegraphics[width=\linewidth]{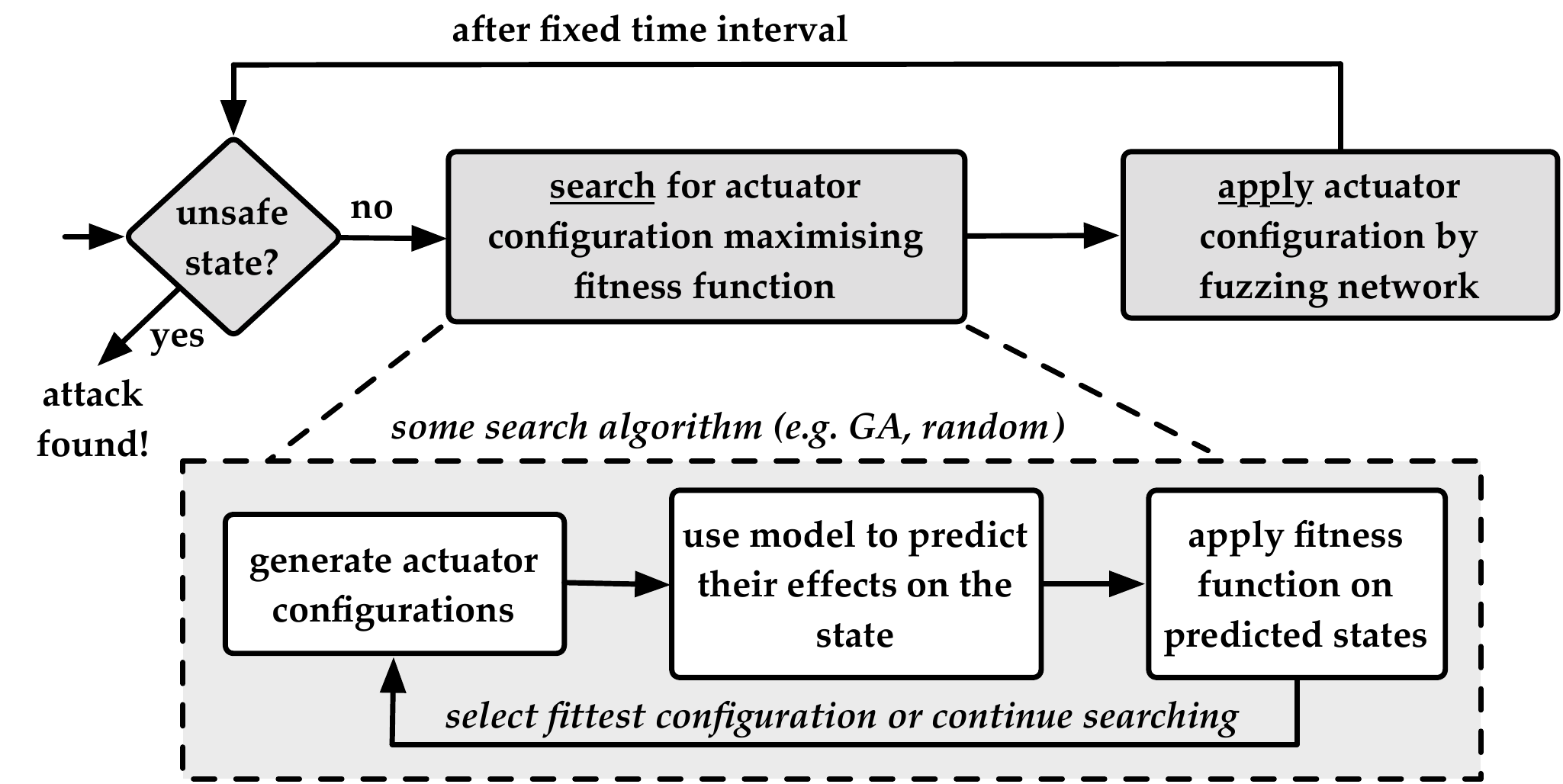}
	\caption{Overview of our ML-guided actuator fuzzing (details of any particular search algorithm are omitted here)}
	\label{fig:overview_of_approach}
\end{figure}

The second step of our approach searches for commands to fuzz the actuators with that will drive the CPS into an unsafe physical state. The sub-steps of this part are summarised in Figure~\ref{fig:overview_of_approach}. To find the right commands, our approach applies a search algorithm over the space of actuator configurations, returning the configuration that is predicted by the model (of the first step) to drive the CPS the closest to an unsafe state. We explore different search algorithms for this task, including random, but also metaheuristic (e.g.~genetic algorithms) given that the state space of actuators can grow quite large (e.g.~$2^{26}$ possible configurations in \swat). We use \emph{fitness functions} to evaluate predicted sensor states with respect to the attack goal.

\substepseparator

\noindent\textbf{Assumptions.} Given that smart fuzzing, ultimately, is intended to generate attacks for evaluating CPS defence solutions, it is important to detail our assumptions about systems and attackers, thus characterising the kinds of attacks that will be found (and just as importantly, those that will not be).

In the learning phase of our approach, we assume the total integrity of the data that the machine learning models are trained on (i.e.~an attacker has not manipulated the learning of the model). In the fuzzing phase, we search for network attacks that could be executed by attackers that are able to compromise the communication links between PLCs and actuators. In particular, we assume that attackers can monitor all genuine sensor readings and actuator commands, and can also inject arbitrary actuator commands at will, controlling their operation independently of any PLC. Furthermore, we assume that genuine commands originating from PLCs can be intercepted and blocked from ever reaching the actuators. Note that while in general we assume a rather powerful attacker, it is possible to use our technique with \emph{more restricted capabilities} too (e.g.~particular actuators or stages only), in order to test defence mechanisms in a more realistic context (see e.g.~Section~\ref{sec:experiments_and_discussion}, RQ5). Attacks of this kind have been reported in practice, e.g.~the discovery that infiltrators had ``manipulated the valves controlling the flow of chemicals'' in an industrial water treatment plant~\cite{Leyden16a}. Furthermore, our focus on compromised actuators can lead to classes of attacks difficult for control theoretic defences to detect, e.g.~\cite{Choi-et_al18a}.

While our approach searches for attacks based on manipulating actuators, it is possible to implement identified attacks in more indirect ways too. For example, instead of spoofing actuator commands over the network, one could spoof a sensor reading so that the system will issue the desired command to the actuator itself. One could also modify some PLC code to cause it to issue the same commands~\cite{Klick-et_al15a,Abbasi-Hashemi16a,Garcia-et_al17a} (thus testing any code attestation mechanisms), or could have a user physically interfering with the plant equipment (i.e.~testing for insider threats).


\subsection{Step One: Learning a Prediction Model}
\label{sec:learning_model}


\noindent\textbf{Data Collection.} Our method requires a dataset from which the relationship between actuators and the evolution of the physical state can be learnt. A suitable raw format for such data is a \emph{time series} of sensor readings and actuator configurations, recorded at regular intervals during the regular operation of the CPS. The required size of the time series depends upon how many `modes' of behaviour the system exhibits, and how quickly the effects of actuator commands are propagated through the physical state. In general, logs from several days of operation may be sufficient to span enough of the system's normal operational behaviour.

Since our method is aimed at engineers or researchers who are seeking to assess the defences of a CPS (i.e.~not an external attacker), we assume that such data is readily available, e.g.~from the system's historian. But in principle, it could also be collected by monitoring the network traffic. The data could also be augmented by injecting manipulations at this stage, so their effects can be incorporated into the learning. We choose not to do this, however, so as to avoid biasing our approach towards pre-determined attacks.

For \swat and \wadi, time series datasets are already available online~\cite{Goh-et_al16a}, and were obtained by running the testbeds non-stop for seven (resp.~14) days without any interruption from attacks or faults. The datasets include the states of all sensors and actuators as recorded by the historians every 1s.

\substepseparator

\noindent\textbf{Training a Model.} Our method requires a model with the input/output relation of Figure~\ref{fig:model_input_output}. As long as the predictions have a high degree of accuracy, however, the particular choice of algorithm for learning that model is immaterial. In this paper, we consider two different ML algorithms that are popular and well-suited for the particular form of training data---a time series of actuator states and (continuous) sensor values. First, we consider a Long Short-Term Memory~(LSTM) network~\cite{Gers-Schmidhuber-Cummins00a}, a deep learning model with an architecture that supports the learning of longer-term dependencies in the data. Second, we consider a Support Vector Regression~(SVR) model~\cite{Drucker-et_al96a}, an extension of support vector machines for handling continuous values (instead of classification).

To apply these ML algorithms to a dataset, the sensor and actuator values must be extracted from the raw logs into a fixed vector format. As we want to learn the relation between actuator states and the evolution of sensor values, a possible form of the vector is $\langle \langle s_0, s_1, s_2, \dots , a_0, a_1, a_2, \dots \rangle, \langle s_0', s_1', s_2', \dots \rangle \rangle $, where the $s_i$s and $a_i$s are the readings/states of sensors and actuators at time point $t$, and the $s_i'$s are the readings of the same sensors but at time point $t+i$ for some fixed time interval~$i$. Depending on the particular algorithm, the sensor values may also need to be normalised. After extracting as many vectors as possible from the raw data, as is typical in learning, the algorithm should only be trained on a fixed portion of them (e.g.~$80\%$), leaving the remainder as test data for assessing the accuracy of the learnt model's predictions.

For \swat and \wadi, we trained two sets of LSTM and SVR models using standard Python library implementations: the Keras neural networks API~\cite{Chollet-et_al15a} and scikit-learn~\cite{Buitinck-et_al13a} respectively. We extracted vectors for training from days 2--5 of the \swat dataset and from days 5--10 of \wadi's, with vectors from the remaining days reserved for testing the learnt models. Our two LSTM models used a traditional architecture consisting of an LSTM layer followed by a dense, fully-connected layer, and took approximately two days of training on a GTX 1070 Ti GPU for each testbed. The accuracy of both models was measured as $97\%$ (with a tolerance between the actual and predicted values of $5\%$). Our two SVR models took approximately half a day each to train, and upon testing, their accuracy was measured as $94\%$ for \swat and $97\%$ for \wadi (again, with a tolerance of $5\%$).

We remark that in our \swat/\wadi implementations, in addition to learning models able to predict \emph{all} the sensor values $\langle s_0', s_1', s_2', \dots \rangle$, we also learnt a series of simpler models that predict \emph{only} $\langle s_0' \rangle$, or $\langle s_1' \rangle$, or $\langle s_2'\rangle$, or \dots. These models are useful in later experiments (Section~\ref{sec:evaluation}) when the fuzzer is attempting to drive one sensor in particular to an unsafe state, since they can provide the necessary predictions more efficiently, and are also faster to learn (5-10 minutes on an off-the-shelf laptop). Furthermore, for SVR, it allowed us to vary the time interval in predictions for different sensors. While most were set at 1s (as for LSTM), we found that we needed a larger interval of 100s in SVR for the water tank level sensors, due to the fact that tank levels change more slowly. (This is only an issue for SVR, since the architecture of LSTM is specially designed to handle long lags between events.)

\subsection{Step Two: Fuzzing to Find Attacks}
\label{sec:optimising}




\noindent\textbf{Fitness Functions.} Our approach uses \emph{fitness functions} to quantify how close some (predicted) sensor readings are to an unsafe state we wish to reach. Intuitively, a fitness function takes a vector of sensor values as input, and returns a number that is larger the `closer' the input is to an unsafe state. The goal of a search algorithm is then to find actuator configurations that are predicted (by the LSTM/SVR models) to bring about sensor states that \emph{maximise} the fitness function.

Fitness functions are manually defined for the CPS under consideration, and should characterise \emph{precisely} what an unsafe state is. There is a great deal of flexibility in how they are defined: it is possible to define them in terms of the unsafe ranges of individual sensors, of groups of related sensors, or of different combinations therein. In this paper, we perform smart fuzzing using fitness functions for the individual sensors in turn, to ensure a suitable variety of attacks, and to avoid the problem of a single sensor always dominating (e.g.~because it is easier to attack).

Defining a fitness function for a single sensor $s$ is straightforward. If its unsafe range consists of all the values above a certain threshold, then a suitable fitness function would be the sensor reading $s$ itself, since maximising it brings it closer to (or beyond) that threshold. If its unsafe range consists of values below a certain threshold, then a suitable fitness function would be the \emph{negation} of the sensor reading, $-s$.

A general fitness function can be defined for any group of the system's sensors. Let $v_s$ denote the current value of sensor $s$, $L_s$ denote its lower safety threshold, $H_s$ denote its upper safety threshold, and $r_s = H_s - L_s$ denote its range of safe values. Let
\[ d_s = \left\{
        \begin{array}{ll}
            \mathrm{min}\left( \left| v_s - L_s \right|, \left| v_s - H_s \right| \right) & \quad L_s \leq v_s \leq H_s \\
            0 & \quad \mathrm{otherwise}
        \end{array}
    \right. \]

\noindent i.e.~the absolute distance of the sensor reading from the nearest safety threshold. Then, the fitness function for a set $S$ of the CPS's sensors is $ \sum_{s \in S} \frac{1}{d_s / r_s} $. This function tends to infinity as individual sensor readings get closer to either their lower or upper safety thresholds. In other words, the fitness function returns an increasingly large number as the system moves in the direction of an unsafe state.

For \swat and \wadi, we defined fitness functions on a per-sensor basis, and treated the values above and below these thresholds as two \emph{separate} cases. We did this for two reasons: first, to ensure that smart fuzzing can find attacks that violate each sensor in both ways (where applicable); and second, as it increases the diversity of attacks in cases where it is much easier to attack a sensor in one direction. We remark that a preliminary investigation found more general fitness functions (e.g.~favouring \emph{any} kind of attack) not to be useful for \swat or \wadi, as the attacks found by smart fuzzing were dominated by sensors that were easier to drive to unsafe ranges, such as the water flow indicators in the first stage.

\substepseparator

\noindent\textbf{Searching and Fuzzing.} We consider two different algorithms for searching across the space of actuator configurations. First, we consider a simple random search, in which several configurations of actuators are randomly generated. The network is then fuzzed to apply the configuration that is predicted (by the LSTM/SVR model) to maximise the fitness function.

\begin{algorithm}[!t]
\caption{GA for Actuator Configurations}\label{genetic_algorithm}
\small
\KwIn{Vector of sensor readings $S$, prediction model $M$, fitness function $f$, population size $n$, no.~of parents $k$, mutation probability $p_m$}
\KwOut{Vector of actuator configurations}
Randomly generate population $P$ of $n$ actuator configurations; \\
Compute fitness of each candidate $c\in P$ with $f(M(S,c))$;\\
\Repeat{timeout}
{
Select $k$ parents from $P$ using Roulette Wheel Selection; \\
Generate new candidates from parents using crossover; \\
Generate new candidates from parents using bit flip mutation with probability $p_m$; \\
    Compute fitness of new candidates $c$ with $f(M(S,c))$; \\
    Replace $P$ with the $n$ fittest of the new and old candidates; \\
}
\Return candidate $c\in P$ that maximises $f(M(S,c))$;
\end{algorithm}

Second, we consider a genetic algorithm~(GA)~\cite{Goldberg89a}, a metaheuristic search inspired by the ``survival of the fittest'' principle from the theory of natural selection. The high-level steps of our particular GA for finding actuator configurations are summarised in Algorithm~\ref{genetic_algorithm}. First, a population of actuator configurations is randomly generated. We assume in this paper that the states of actuators are discrete, allowing us to encode each solution as a simple bit vector (e.g.~$\mathtt{0001100011001}\dots$). Most actuators can only be in one of two states and thus need only one bit in the vector (i.e.~$\mathtt{0}$ for off, $\mathtt{1}$ for on), though some actuators (e.g.~modulating control valves) have more states and need additional bits in the encoding. Next, we calculate the fitness of each candidate in the population by: (1)~applying the LSTM/SVR model; then (2)~applying the fitness function to the resulting predicted state.

At this point we enter the main loop, in which the fittest candidates are selected for generating ``offspring'' from. We select the candidates using roulette wheel selection~\cite{Goldberg89a}, which assigns to each candidate a probability of being selected based on its fitness. If $f_i$ is the fitness of one of the $n$ candidates, then its probability of being selected is $ f_i / \sum_{j=1}^n f_j$. Next, we sample candidates based on the probabilities using the following implementation. First, a random number is generated between 0 and the sum of the candidates' fitness scores. Then, we iterate through the population, until the accumulated fitness score is larger than that number. At this point we stop, selecting the last candidate as a ``parent''. We repeat this until we have selected $k$ parents (possibly including duplicates).

From the parents, we generate new candidates (offspring)  by applying crossover and mutation. In the former, we randomly choose a point in the bit vectors of two parents, and swap the sub-vectors to the right of it. In the latter, we randomly flip a bit (from $\mathtt{0}$ to $\mathtt{1}$ or vice versa) with some fixed probability, $p_m$. The fitness of all the new candidates is calculated, then the $n$ fittest candidates from the old and the new candidates are carried forward as the new population, with the other, less fit candidates all eliminated. This iteration continues until a fixed timeout is reached. The fittest actuator configuration is then returned, and the relevant commands to apply it are issued.

For \swat, implementing the search algorithms was relatively straightforward, especially as its actuators consist only of binary states (on or off) and can be encoded directly as bit vectors. For \wadi the same is true, except for its modulating control valve feeds: these are set to an integer between 0 and 100, so we encoded their states as 7 bit binary numbers. For both the random and GA searches, we set a timeout of 10s, after which the best actuator configuration found so far would be the one that is applied. While this 10s timeout is clearly longer than the 1s prediction time interval of our models (except for the 100s time interval for tank levels in SVR), the predictions are still meaningful because the physical states of the testbeds evolve so slowly in the meantime. For CPSs that change faster, a longer prediction time interval may need to be considered for the search to remain practically useful.

In our GA implementation, we set the population size as $n=100$, number of parents as $k=100$, and the mutation probability as $0.1$. These parameters were chosen to ensure that the algorithm will converge before the 10s timeout. For \wadi, any mutation that would change a modulating control valve's value to something outside of its range is rejected. Furthermore, we cap the maximum number of iterations at 100 (if reached before the timeout).

Once our random or GA search algorithm identifies the fittest actuator configuration, we use the Python package pycomm~\cite{Ruscito} to fuzz the actuators over the network. As long as the system continues to approach the targeted unsafe state, the actuator configuration is held; otherwise, the search is repeated to identify a more fruitful configuration.

\section{Evaluation} \label{sec:evaluation}

We evaluate the effectiveness of smart fuzzing on the \swat and \wadi testbeds (Section~\ref{sec:swat_testbed}). 

\subsection{Research Questions}\label{sec:research_questions}

Our evaluation addresses five research questions based on our original design requirements for smart fuzzing~(Section~\ref{sec:introduction}):

\begin{description}
	\item[\textbf{RQ1 (Efficiency):}] \hspace{48pt} How quickly is smart fuzzing able to find a targeted attack?
	\item[\textbf{RQ2 (Comprehensiveness):}] \hspace{95pt}How many unsafe states can the attacks of smart fuzzing cover?
	\item[\textbf{RQ3 (Setup):}] \hspace{36pt}Which combinations of model and search algorithm are most effective?
	\item[\textbf{RQ4 (Comparisons):}] \hspace{65pt} How do the attacks compare against those of other approaches or those in benchmarks?
	\item[\textbf{RQ5 (Utility):}] \hspace{36pt}Are the discovered attacks useful for testing CPS attack detection mechanisms?
\end{description}

\noindent RQs~1--2 consider whether smart fuzzing achieves its principal goal of finding network attacks. We assess this from two different angles: first, in terms of how quickly it is able to drive the CPS into a particular unsafe state; and second, in terms of how many different unsafe states the attacks can cover. RQ~3 considers how different setups of smart fuzzing (i.e.~different models or search algorithms) impact its ability to find attacks. RQ~4 compares the effectiveness of smart fuzzing against other approaches: first, the baseline of randomly mutating actuator states \emph{without} reference to a model of the system; and second, an established, manually constructed benchmark of attacks~\cite{Goh-et_al16a}. Finally, RQ~5 investigates whether the attacks found by smart fuzzing are useful for testing existing cybersecurity defence mechanisms. We assess this final question using \swat, as practical defence mechanisms are more established for this testbed; in particular, its ``Water Defense'' invariant-based detection solution~\cite{Adepu-Mathur16c,Adepu-Mathur18b}.

\subsection{Experiments and Discussion}\label{sec:experiments_and_discussion}

We designed three experiments for the \swat and \wadi testbeds to evaluate our research questions. The programs we built to perform these experiments and all supplementary material are available online to download~\cite{Supplementary-Material}.

\substepseparator

\noindent\textbf{Experiment \#1: RQs 1--3.} In our first experiment, we systematically target the different possible unsafe sensor ranges in \swat and \wadi, using our tools in four different model/search setups: LSTM-GA, LSTM-Random, SVR-GA, and SVR-Random (see Section~\ref{sec:implementing_smart_fuzzing}). Our goal is to collate the data and obtain an overall picture of how the different ML-guided setups perform, how quickly targeted attacks are found, and how many different unsafe states are covered.

For each of the setups and fitness functions (usually two per sensor, targeting the lower and upper safety thresholds separately), we ran the experiment as follows. First, we ``reset'' the testbed by operating it normally until all sensors enter their safe ranges. Upon reaching this state, we launch the fuzzer with the given setup and fitness function, and let it run without interference for up to 20 minutes (long enough to ensure that over/underflow attacks are able to complete). If the unsafe state targeted by the fitness function is reached within that time, we disable the fuzzer and record the time point at which the transition to unsafe state occurred. (If other sensors happen to enter their unsafe states along the way, we record this too, but do not disable the fuzzer for them.) Once the fuzzer is disabled, the testbed is allowed to reset and return to a safe state. We repeat these steps 10 times for \emph{every} combination of fuzzer setup and fitness function, and record the median so as to remove biases resulting from differences in starting states.

For \wadi, we note that we use slightly different fitness functions for the flow sensors 2\_FQ\_101, 2\_FQ\_201, \dots, 2\_FQ\_601 which monitor the supply of the testbed's consumer tanks. Here, the goal is not simply to disrupt the supply of a consumer (e.g.~maximising $-s$ for the targeted sensor $s$), but rather to do so without disrupting the supplies of any others. If targeting the first consumer tank, for example, this goal is expressed by the fitness function: $ \frac{1}{\mathrm{2\_FQ\_101}} \ast \mathrm{2\_FQ\_201} \ast \mathrm{2\_FQ\_301} \ast \cdots \ast \mathrm{2\_FQ\_601}$, which is maximised when the supply of the first consumer tank is cut off, while remaining sensitive to any possible (unwanted) effects on the other tanks.

\begin{table*}[!t]
	\renewcommand{\arraystretch}{1.3}
\caption{Results: median time taken (s) to drive \swat's Flow, Differential Pressure, and Level Indicator Transmitters into unsafe states}
\label{tab:results_flow}
	\centering
\begin{tabular}{|c|c||ccc|ccccc|c|ccc|ccc|}
	\cline{3-17}
	\multicolumn{2}{c|}{} &\multicolumn{3}{c|}{Flow (High)} &\multicolumn{5}{c|}{Flow (Low)} & Pr.~(L) & \multicolumn{3}{c|}{Tanks (Overflow)} & \multicolumn{3}{c|}{Tanks (Underflow)} \\ 
	\cline{3-17}
	\multicolumn{2}{c|}{} &  \rotatebox{90}{FIT101} & \rotatebox{90}{FIT201} & \rotatebox{90}{FIT601} &  \rotatebox{90}{FIT101} & \rotatebox{90}{FIT201} & \rotatebox{90}{FIT301} & \rotatebox{90}{FIT401} & \rotatebox{90}{FIT501} & \rotatebox{90}{\scriptsize DPIT301} & \rotatebox{90}{LIT101} & \rotatebox{90}{LIT301} & \rotatebox{90}{LIT401} & \rotatebox{90}{LIT101} & \rotatebox{90}{LIT301} & \rotatebox{90}{LIT401} \\ 
	\hhline{--===============}
	\parbox[t]{2mm}{\multirow{4}{*}{\rotatebox[origin=c]{90}{ML-Based}}} & LSTM-GA &  14 & 17 & 20 & 14 & 7 & 17 & 16 & 5  & 12 & 333 & \textbf{496} & 729 &   511 & 1200+ & 1200+     \\
	& LSTM-Random &   15 & 18 & 21$^\ast$ &  14 & 8 & 19 & 14 & 5  & 14 & 339 & 618 & 697 &  526 & 1200+ & 1200+      \\
	&SVR-GA &  13 & 10 & 16 &  13 & 7 & 16 & 16 & 5  & 11 & 385 & \textbf{439} & 696 &  569 & 1200+ & 1200+       \\
	&SVR-Random &  15 & 12 & 21 &  14 & 7 & 19 & 16 & 5  & 12 & 337 & 577 & 681 &  655 & 1200+ & 1200+     \\
	\hline
	\parbox[t]{2mm}{\multirow{2}{*}{\rotatebox[origin=c]{90}{Other}}} & Random (No Model) & 15  & 20  & --- & 15 & --- & --- & --- & 5  & --- & 435 & --- & --- & --- & --- & ---      \\
	& Benchmark~\cite{Goh-et_al16a} & 14 & 18 & ---  & 14 & 18 & --- & --- & --- & --- & 454$^\ast$ & 771$^\ast$ & 1200+ & --- & --- & 1200+    \\
	\hline
\end{tabular}
\end{table*}

\begin{table*}[!t]
	\renewcommand{\arraystretch}{1.3}
\caption{Results: median time taken (s) to drive WADI into unsafe flow/level states, and to cut the supply of specific consumers}
\label{tab:results_wadi_1}
	\centering
\begin{tabular}{|c|c||c|cccc|ccc|ccc|cccccc|}
	\cline{3-19}
	\multicolumn{2}{c|}{} &\multicolumn{1}{c|}{Fl.~(H)} & \multicolumn{4}{c|}{Flow (Low)} & \multicolumn{3}{c|}{Tanks (Overflow)} & \multicolumn{3}{c|}{Tanks (Underflow)} & \multicolumn{6}{c|}{Cut Supply of Specific Consumer} \\
	\cline{3-19}
	 \multicolumn{2}{c|}{} & \rotatebox{90}{\scriptsize 3\_FIT\_001}  &  \rotatebox{90}{\scriptsize 1\_FIT\_001} & \rotatebox{90}{\scriptsize 2\_FIT\_001} & \rotatebox{90}{\scriptsize 2\_FIT\_002} &  \rotatebox{90}{\scriptsize 3\_FIT\_001} & \rotatebox{90}{\scriptsize 1\_LT\_001}  &  \rotatebox{90}{\scriptsize 2\_LT\_002} & \rotatebox{90}{\scriptsize 3\_LT\_001} & \rotatebox{90}{\scriptsize 1\_LT\_001} &  \rotatebox{90}{\scriptsize 2\_LT\_002} & \rotatebox{90}{\scriptsize 3\_LT\_001} & \rotatebox{90}{\scriptsize 2\_FQ\_101}  &  \rotatebox{90}{\scriptsize 2\_FQ\_201} & \rotatebox{90}{\scriptsize 2\_FQ\_301} & \rotatebox{90}{\scriptsize 2\_FQ\_401} &  \rotatebox{90}{\scriptsize 2\_FQ\_501} & \rotatebox{90}{\scriptsize 2\_FQ\_601}  \\
	\hhline{--=================}
	\parbox[t]{2mm}{\multirow{4}{*}{\rotatebox[origin=c]{90}{ML-Based}}} & LSTM-GA & 9 & 7  & 7 & 9 & 8  & 1200+ & 879 & 1200+ & 801  & 1200+   & $746^\dag$   & 15 & 15  & 15 & 15 & 15   &   15    \\   
	& LSTM-Random & 8 & 7  & 7 & 9 & 8    & 1200+ & 744  & 1200+  & 799 & 1200+  & $746^\dag$  & --- & ---  & --- & --- & ---  &   ---    \\
	& SVR-GA & 8 & 7  & 7  & 9 & 9  & 1200+ & 841  & 1200+  & 731 & 1200+  & $746$  & 15 & 15  & 15 & 16 & 15  &   16    \\
	& SVR-Random & 9 & 7  & 7  & 9  & 9   & 1200+ & 904  & 1200+  & 767  & 1200+  & $746^\dag$   & --- & ---  & --- & --- & ---  &  ---   \\
	\hline
	\hspace{-1mm}{\small Oth.}\hspace{-1mm} & Random (No Model) & 8 & 7  & 7 & --- &  8  & --- & --- & --- & --- & --- & ---      & --- & --- & --- & --- & --- & ---       \\
	\hline
\end{tabular}
\end{table*}


\emph{Results.} The results of this experiment are given in the first four rows of Tables~\ref{tab:results_flow} and \ref{tab:results_wadi_1}. The rows denote the different smart fuzzing setups, whereas the columns denote the different possible unsafe states that were targeted for each sensor. These include: Flow Indicator Transmitters~(FITs/FQs), which measure water flow in pipes and can become too High or Low; a Differential Pressure Sensor~(DPIT), which can become too High or Low; and Level Indicator Transmitters~(LITs/LTs), which measure the water levels of tanks, and can indicate a risk of Overflow or Underflow. We do not include the Analyser Indicator Transmitters~(AITs), which measure properties such as pH, as the testbeds currently take raw water from the mains (i.e.~already close to pH 7); as a result, the readings barely vary during the plant's operation. The numbers indicate the amount of time taken, in seconds, for a particular fuzzing setup to reach a targeted unsafe state; they are the medians obtained from 10 repetitions per combination of fuzzing setup and fitness function. Numbers indicated with an asterisk ($\ast$) indicate that one or more repetitions that were unable to reach the unsafe state within 20 minutes. Furthermore, $1200+$ indicates that despite approaching the given unsafe state, none of the repetitions were able to cross the threshold.

For most of the targeted unsafe states, the performance~\textbf{(RQ1, RQ3)} of smart fuzzing across the four different setups is essentially the same. This suggests that for a system of the testbeds' complexity, having \emph{some} reasonably accurate prediction model and a lightweight search algorithm (e.g.~random) is sufficient to identify several attacks. 

\begin{center}
\noindent\fbox{%
    \parbox{0.75\linewidth}{%
        \small\emph{Attacks covering most of the individual sensors can be discovered using an ML model together with a simple search algorithm (e.g.~random).}
    }
}
\end{center}

\noindent There is one exception to report: the LSTM-GA and SVR-GA setups noticeably outperformed their Random variants at driving \swat's water tank level sensor LIT301 into an unsafe state for overflow (Table~\ref{tab:results_flow}). In comparison to other sensors, there are fewer configurations of actuators that can drive LIT301 into its overflow range (since the sensor is relevant to four of the six stages); they are harder to find by ``accident'' (as in a random search), which we believe is why the fuzzing setups applying metaheuristic search were more performant.

The four fuzzing setups were able to cover the same number of unsafe states~\textbf{(RQ2, RQ3)}, but with one exception: when aiming to cut the supply of specific consumer tanks in \wadi (Table~\ref{tab:results_wadi_1}), we found that the ML+Random setups failed on all counts, whereas ML+GA could succeed across all sensors.

\begin{center}
\noindent\fbox{%
    \parbox{0.75\linewidth}{%
        \small\emph{If attacks can only occur under strict conditions, then a more sophisticated search algorithm (e.g.~GA) is necessary.}
    }
}
\end{center}

\wadi's tank level sensor 3\_LT\_001 (\dag) is a special case in Table~\ref{tab:results_wadi_1}: all four setups can generate the actuator commands necessary to cause an underflow. However, it takes too long to repeat the experiments, as once drained, it takes a \emph{very} long time for the tank in this phase to return to its normal range. Our experiments were performed for SVR-GA only, but since the other setups all found the right actuator configuration immediately, we would expect to see similar results.

For some tank level sensors, smart fuzzing approached (but did not cross) the over/underflow thresholds. One reason might be that during the learning stage, not enough data was observed that was relevant to the evolution of those sensors, leading to a poorer prediction model. In future work, we will address this by either fuzzing the actuators during the learning stage (so that the model can train on some additional, abnormal behaviours), or by using an online learning scheme in which the model is updated over time as more data is observed. Some sensed properties were not possible to attack at all and are excluded from the table, e.g.~the pressure sensor PIT501 which was suffering a hardware fault at the time of experimentation.

\substepseparator

\noindent\textbf{Experiment \#2: RQ 4.} Our second experiment seeks to compare the results of our smart fuzzing setups against those of other approaches. As we are unaware of any existing tools we can make a direct comparison with, we instead make a comparison against two different baselines. First, we compare against an automatic approach in which all of the ``smartness'' is removed, and commands to the actuators are simply generated \emph{and applied} randomly, without reference to any prediction model for the testbed (in contrast to LSTM-Random and SVR-Random). Second, and in contrast, we compare against the attacks of an established, expert-crafted benchmark~\cite{Goh-et_al16a} (for \swat only), which were systematically derived from an attack model. For the different attacks derived from these two sources, this experiment is roughly similar to the first: launch them, and record which sensors are driven into unsafe ranges (and at which time points). The idea of the experiment is to establish where the effectiveness of smart fuzzing can be positioned between two extremes: a simplistic, uninformed search on the one hand; and an expert-crafted, comprehensive benchmark on the other.

We ran the experiment as follows. For the random baseline, we wrote a program to randomly generate 10 distinct sets of actuator configurations. For each set in turn, we begin by `resetting' \swat to a normal state, using the same procedure as the previous experiment. Once in such a state, we then fuzzed the network to override the actuator states with the given random configuration for 20 minutes. If during the run any sensor readings were driven into an unsafe range, we recorded the sensors (and ranges) in question, and the time points at which they \emph{first} became unsafe. After the run, the system was allowed to reset, before repeating the process for the other randomly generated configurations.

For the benchmark~\cite{Goh-et_al16a}, we manually extracted all attack sequences that were intended drive the system into unsafe physical states. For each of these six attacks in turn, we fuzzed the network manually to recreate the attack sequence, overriding the sensor and actuator states as prescribed by the benchmark. If during the run any (\emph{actual}) sensor readings were driven into an unsafe range, we recorded the sensors and time points at which they \emph{first} became unsafe. This was repeated 10 times for each attack sequence.


\emph{Results.} The results of this experiment are given in the bottom rows of Tables~\ref{tab:results_flow}--\ref{tab:results_wadi_1}. The numbers indicate the time at which sensor entered an unsafe state for the first time. For ``Random (No Model)'', they are the medians across runs for 10 randomly generated actuator configurations; for ``Benchmark'', they are the medians across 10 repetitions for each of the six attacks that target some unsafe state. The dashes (---) indicate which sensors never reached an unsafe state.

In comparing these numbers with smart fuzzing~\textbf{(RQ4)}, it appears that our approach drives \emph{more} of the sensors to unsafe states and does so \emph{faster}. The comparison with Random (No Model) makes clear that most of the attacks are not possible to find by ``accident'', or just by simply fuzzing at random without any artificial or human intelligence. At the other end, the comparison with the \swat benchmark illustrates that smart fuzzing was able to drive the system to six additional types of unsafe states that were not covered by the benchmark. Having said that, the benchmark does include several attacks of the kind we do not consider (e.g.~manipulating a sensor or actuator, but without the goal of reaching an unsafe state); and while its comparable attacks were slower than those of smart fuzzing, efficiency was not necessarily a goal. The comparison does however suggest that our approach could complement the benchmark and improve its comprehensiveness.

\begin{center}
\noindent\fbox{%
    \parbox{0.75\linewidth}{%
        \small\emph{Smart fuzzing complemented an established benchmark by covering six additional unsafe states.}
    }
}
\end{center}

\substepseparator

\noindent\textbf{Experiment \#3: RQ 5.} Our final experiment explores, by means of a case study, whether smart fuzzing may have some utility for testing the attack defence mechanisms of CPSs. As these mechanisms are less established for \wadi (e.g.~forced logic between actuators, which our method could easily overcome), we considered \swat's ``Water Defense''~(WD) solution~\cite{Adepu-Mathur16c,Adepu-Mathur18b}, which has demonstrated its efficacy many times in the past (e.g.~detecting attacks at professional \swat hackathons~\cite{Adepu-Mathur18a}). This attack detection mechanism is based on monitoring several different expected conditions and invariants of the system, and highlighting to the operator whenever one of them is violated. Broadly speaking, WD monitors two kinds of properties:~(1)~\emph{state-dependent conditions}, and (2)~\emph{physical invariants}. State-dependent conditions define expected relationships between sensors and actuators, e.g.~``if motorised valve MV101 is open, then the flow indicator FIT101 must be non-zero''~\cite{Adepu-Mathur18a}. Physical invariants are properties that are expected to \emph{always} be true of a system behaving normally, i.e.~they mathematically characterise the physical process.

In this case study, we aimed to systematically drive each sensor in \swat towards an unsafe state \emph{while avoiding detection} by the monitors of WD. For a more fine-grained assessment of the attack detection mechanism, we used it with (1)~all checks enabled; (2)~only state-dependent conditions enabled; and (3)~only physical invariants enabled. Using our SVR-GA setup and each fitness function in turn, we ran our tools until either driving the targeted sensor to an unsafe state, or violating an invariant. This was repeated for all fitness functions and all three modes of the WD system. We remark that in this experiment, the search algorithms were constrained to manipulating \emph{only} the actuators within stage(s) of \swat relevant to the targeted sensor. This was to reduce the possibility of unnecessarily triggering alerts from stages that were not directly relevant to the target of the attack.


\emph{Results.} We found that when enabling all checks in WD, or only the state-dependent conditions, smart fuzzing was unable to successfully attack \swat without detection. This is unsurprising, as the attack detection mechanism is especially effective at detecting process attacks: at the recent \swat hackathon, comprising teams from industry and academia, WD was able to detect all but one of the process attacks involving valve or pump manipulations, and more of the attacks than the industry teams assigned to system defence~\cite{Adepu-Mathur18a}.

However, smart fuzzing was able to successfully attack two sensed properties---FIT501~(Low) and FIT601~(High)---without detection by the \emph{physical invariants}. This weakness in the physical invariants could potentially be exploited by a real attacker if they are able to spoof the sensor and actuator values considered by the state-dependent conditions.

\emph{Attack \#1.} The first attack targeted the flow sensor FIT501 in stage five of \swat (reverse osmosis process). The flow sensor is associated with the two high-pressure pumps, P501 and P502, that pump the water in from stage four. Prior to launching the attack, P501 was on and P502 was in standby.

To achieve the goal of decreasing the flow of FIT501, smart fuzzing automatically generated and launched the following simple attack over the star network of \swat: turn off pump P501; turn off P502. Within about 5 seconds, the flow through the pipes reduces to zero, and FIT501 enters an unsafe state. The physical impact of this attack is a reduction in reverse osmosis production, with possible effects from the reduction cascading through the system.

\emph{Attack \#2.} The second attack targeted the flow sensor FIT601 in stage six of \swat, which is a backwash process that is intrinsically linked with the other stages. It stores purified water from stage five, ready to be recycled back into stage one; it also stores rejected water from stage five, ready to be pumped to clean the ultra filtration of stage three.

To achieve the goal of increasing the value of FIT601, smart fuzzing automatically generated and launched the following attack over the star network of \swat: open the motorised valves MV301, MV302, MV303, MV304; turn on the pumps P601 and P602; and ensure P301 and P302 are in the same state. This actuator configuration increases the flow of FIT601 beyond acceptable levels from the moment the valves are opened. At this stage, there are further physical consequences of the attack, depending on whether smart fuzzing chose to turn both P301 and P302 off or left them both on. In the former case, ultra filtration is stopped completely, but water continues to arrive from the backwash, leading to a reduction of output for the process. In the latter case, ultra filtration is still operating, but the increase in flow from the backwash stage leads to the pressure increasing beyond the normal operating values. From running the fuzzer through to reaching the unsafe state, the attack takes approximately 15-20 seconds.

\begin{center}
\noindent\fbox{%
    \parbox{0.75\linewidth}{%
        \small\emph{Smart fuzzing discovered two attacks that evaded detection by the physical invariant monitors of the established ``Water Defense'' mechanism for \swat.}
    }
}
\end{center}

\subsection{Threats to Validity}

Finally, we remark on some threats to the validity of our evaluation. First, our approach was implemented for CPS \emph{testbeds}: while they are real, fully operational plants based on the designs of industrial ones, they are still smaller, and our results may therefore not scale-up (this is difficult to test due to the confidentiality surrounding plants in cities). Second, the initial states of the testbeds were not controlled, other than to be within their normal ranges, meaning that our performance results may vary slightly. Finally, for testing CPS attack detection mechanisms, we only studied an \emph{invariant-based} solution, meaning that our conclusions may not hold for other types of defences (to be addressed in future work).

\section{Related Work} \label{sec:related_work}

In this section, we highlight a selection of the literature that is particularly relevant to some of the main themes of this paper: constructing attacks, fuzzing, and assessing robustness. We remark that related works on \emph{attack defence mechanisms} are discussed earlier in the paper in Section~\ref{sec:introduction}, with some mechanisms specific to \swat/\wadi discussed in Section~\ref{sec:swat_testbed}.

A number of papers have considered the systematic \emph{construction} or \emph{synthesis} of attacks for CPSs in order to test (or demonstrate flaws in) some specific attack detection mechanisms. Liu et al.~\cite{Liu-et_al11a}, for example, target electric power grids, which are typically monitored based on state estimation, and demonstrate a new class of data injection attacks that can introduce arbitrary errors into certain state variables and evade detection. Huang et al.~\cite{Huang-et_al18a} also target power grids, presenting an algorithm that synthesises attacks that are able to evade detection by conventional monitors. The algorithm, based on ideas from hybrid systems verification and sensitivity analysis, covers both discrete and continuous aspects of the system. Urbina et al.~\cite{Urbina-et_al16a} evaluate several attack detection mechanisms in a comprehensive review, concluding that many of them are not limiting the impact of stealthy attacks (i.e.~from attackers who have knowledge about the system's defences), and suggest ways of mitigating this. Sugumar and Mathur~\cite{Sugumar-Mathur17a} address the problem of assessing attack detection mechanisms based on process invariants: their tool simulates their behaviour when subjected to single stage single point attacks, but must first be provided with some formal timed automata models.

C\'{a}rdenas et al.~\cite{Cardenas-et_al14a} propose a general framework for assessing attack detection mechanisms, but in contrast to the previous works, focus on the business cases between different solutions. For example, they consider the cost-benefit trade-offs and attack threats associated with different methods, e.g.~centralised vs.~distributed.

\emph{Fuzzing} has been a popular research topic in security and software engineering for many years, but the goals of previous works tend to differ from ours, which is a general approach/tool for discovering CPS network attacks. Fuzzing has previously been applied to CPS models, but for the goal of \emph{testing} them, rather than finding attacks. CyFuzz~\cite{Chowdhury-Johnson-Csallner17a} is one such tool, which offers support for testing Simulink models of CPSs. American fuzzy lop~\cite{Zalewski} targets programs, and uses GAs to increase the code coverage of tests and find more bugs. Cha et al.~\cite{Cha-Woo-Brumley15a} also target software, using white-box symbolic analysis on execution traces to maximise the bugs it finds in programs. A number of works (e.g.~\cite{Vigna-et_al04a}) have targeted the fuzzing of network protocols in order to test their intrusion detection systems. In contrast, our work starts from the assumption that an attacker has \emph{already compromised} the network, and uses ML-guided fuzzing to find the different ways that such an attacker might drive the system to an unsafe state. The attacks that it uncovers then form test suites for attack detection mechanisms.

There are more \emph{formal} approaches that could be used to analyse a \cps and construct a benchmark of different attacks. However, these typically require a \emph{formal specification}, which, if available in the first place, may be too simple to capture all the complexities in the physical processes of full-fledged CPSs. Kang et al.~\cite{Kang-et_al16a}, for example, construct a discretised first-order model of \swat's first three stages in Alloy, and analyse it with respect to some safety properties. However, the work uses very simple abstractions of the physical processes, and only partially models the system (we consider \emph{all} of it without the need for any formal model). Castellanos et al.~\cite{Castellanos-Ochoa-Zhou18a} automatically extract models from PLC programs that highlight the interactions among different internal entities of a CPS, and propose reachability algorithms for analysing the dependencies between control programs and physical processes. McLaughlin et al.~\cite{McLaughlin-et_al14a} describe a trusted computing base for verifying safety-critical code on PLCs, with safety violations reported to operators when found. Etigowni et al.~\cite{Etigowni-et_al16a} define a CPS control solution for securing power grids, focusing on information flow analyses based on (potentially verifiable) policy logic and symbolic execution. Beyond these examples, if a CPS can be modelled as a hybrid system, there are several formal methods that can be applied to it, including model checking~\cite{Frehse-et_al11a,Wang-et_al18a}, SMT solving~\cite{Gao-Kong-Clarke13a}, non-standard analysis~\cite{Hasuo-Suenaga12a}, process calculi~\cite{Lanotte-et_al17a}, concolic testing~\cite{Kong-et_al16a}, and theorem proving~\cite{Quesel-et_al16a}. Defining a formal model that accurately characterises enough of the physical process and its interactions with the PLCs is, however, the \emph{hardest} part. Smart fuzzing in contrast can achieve results on real-world CPSs without the need for specifying a model at all: it learns one implicitly and automatically from the data logs.

\section*{Acknowledgements}

We are grateful to the anonymous referees for their feedback on drafts of this paper, as well as to Venkata Reddy for his technical assistance with the WADI testbed. This work was supported in part by the National Research Foundation~(NRF), Prime Minister's Office, Singapore, under its National Cybersecurity R\&D Programme (Award No.~NRF2014NCR-NCR001-040) and administered by the National Cybersecurity R\&D Directorate. Zhang was supported by Alibaba-Zhejiang University Joint Institute of Frontier Technologies, by Zhejiang Key R\&D Plan (Grant No.~2019C03133), and by a Major Scientific Research Project of Zhejiang Lab (Grant No.~2018FD0ZX01). Corresponding authors Poskitt, Sun, and Zhang may be contacted with queries regarding this paper.

\bibliographystyle{IEEEtran}

\bibliography{references}

\begin{thebibliography}{10}
\providecommand{\url}[1]{#1}
\csname url@samestyle\endcsname
\providecommand{\newblock}{\relax}
\providecommand{\bibinfo}[2]{#2}
\providecommand{\BIBentrySTDinterwordspacing}{\spaceskip=0pt\relax}
\providecommand{\BIBentryALTinterwordstretchfactor}{4}
\providecommand{\BIBentryALTinterwordspacing}{\spaceskip=\fontdimen2\font plus
\BIBentryALTinterwordstretchfactor\fontdimen3\font minus
  \fontdimen4\font\relax}
\providecommand{\BIBforeignlanguage}[2]{{%
\expandafter\ifx\csname l@#1\endcsname\relax
\typeout{** WARNING: IEEEtran.bst: No hyphenation pattern has been}%
\typeout{** loaded for the language `#1'. Using the pattern for}%
\typeout{** the default language instead.}%
\else
\language=\csname l@#1\endcsname
\fi
#2}}
\providecommand{\BIBdecl}{\relax}
\BIBdecl

\bibitem{US-NSF18a}
{US National Science Foundation}, ``Cyber-physical systems ({CPS}),''
  \url{https://www.nsf.gov/publications/pub_summ.jsp?ods_key=nsf18538&org=NSF},
  2018, document number: nsf18538.

\bibitem{Rajkumar-et_al10a}
R.~Rajkumar, I.~Lee, L.~Sha, and J.~A. Stankovic, ``Cyber-physical systems: the
  next computing revolution,'' in \emph{Proc.\ Design Automation Conference
  (DAC 2010)}.\hskip 1em plus 0.5em minus 0.4em\relax {ACM}, 2010, pp.
  731--736.

\bibitem{Leyden16a}
\BIBentryALTinterwordspacing
J.~Leyden, ``Water treatment plant hacked, chemical mix changed for tap
  supplies,'' \emph{The Register}, 2016, acc.: September\ 2019. [Online].
  Available:
  \url{https://www.theregister.co.uk/2016/03/24/water_utility_hacked/}
\BIBentrySTDinterwordspacing

\bibitem{ICS-Cert-Alert16a}
{ICS-CERT Alert}, ``Cyber-attack against {Ukrainian} critical infrastructure,''
  \url{https://ics-cert.us-cert.gov/alerts/IR-ALERT-H-16-056-01}, 2016,
  document number: {IR-ALERT-H-16-056-01}.

\bibitem{Cheng-Tian-Yao17a}
L.~Cheng, K.~Tian, and D.~D. Yao, ``{Orpheus}: Enforcing cyber-physical
  execution semantics to defend against data-oriented attacks,'' in
  \emph{Proc.\ Annual Computer Security Applications Conference (ACSAC
  2017)}.\hskip 1em plus 0.5em minus 0.4em\relax {ACM}, 2017, pp. 315--326.

\bibitem{Harada-et_al17a}
Y.~Harada, Y.~Yamagata, O.~Mizuno, and E.~Choi, ``Log-based anomaly detection
  of {CPS} using a statistical method,'' in \emph{Proc.\ International Workshop
  on Empirical Software Engineering in Practice (IWESEP 2017)}.\hskip 1em plus
  0.5em minus 0.4em\relax {IEEE}, 2017, pp. 1--6.

\bibitem{Inoue-et_al17a}
J.~Inoue, Y.~Yamagata, Y.~Chen, C.~M. Poskitt, and J.~Sun, ``Anomaly detection
  for a water treatment system using unsupervised machine learning,'' in
  \emph{Proc.\ {IEEE} International Conference on Data Mining Workshops (ICDMW
  2017): Data Mining for Cyberphysical and Industrial Systems (DMCIS
  2017)}.\hskip 1em plus 0.5em minus 0.4em\relax IEEE, 2017, pp. 1058--1065.

\bibitem{Pasqualetti-Dorfler-Bullo11a}
F.~Pasqualetti, F.~Dorfler, and F.~Bullo, ``{Cyber-physical attacks in power
  networks: Models, fundamental limitations and monitor design},'' in
  \emph{Proc.\ {IEEE} Conference on Decision and Control and European Control
  Conference (CDC-ECC 2011)}.\hskip 1em plus 0.5em minus 0.4em\relax {IEEE},
  2011, pp. 2195--2201.

\bibitem{Aggarwal-et_al18a}
E.~Aggarwal, M.~Karimibiuki, K.~Pattabiraman, and A.~Ivanov, ``{CORGIDS:} {A}
  correlation-based generic intrusion detection system,'' in \emph{Proc.\
  Workshop on Cyber-Physical Systems Security and PrivaCy (CPS-SPC
  2018)}.\hskip 1em plus 0.5em minus 0.4em\relax {ACM}, 2018, pp. 24--35.

\bibitem{Aoudi-et_al18a}
W.~Aoudi, M.~Iturbe, and M.~Almgren, ``Truth will out: Departure-based
  process-level detection of stealthy attacks on control systems,'' in
  \emph{Proc.\ {ACM} {SIGSAC} Conference on Computer and Communications
  Security (CCS 2018)}.\hskip 1em plus 0.5em minus 0.4em\relax {ACM}, 2018, pp.
  817--831.

\bibitem{He-et_al18a}
Z.~He, A.~Raghavan, S.~M. Chai, and R.~B. Lee, ``Detecting zero-day controller
  hijacking attacks on the power-grid with enhanced deep learning,''
  \emph{CoRR}, vol. abs/1806.06496, 2018.

\bibitem{Kravchik-Shabtai18a}
M.~Kravchik and A.~Shabtai, ``Detecting cyber attacks in industrial control
  systems using convolutional neural networks,'' in \emph{Proc.\ Workshop on
  Cyber-Physical Systems Security and PrivaCy (CPS-SPC 2018)}.\hskip 1em plus
  0.5em minus 0.4em\relax {ACM}, 2018, pp. 72--83.

\bibitem{Lin-et_al18a}
Q.~Lin, S.~Adepu, S.~Verwer, and A.~Mathur, ``{TABOR:} {A} graphical
  model-based approach for anomaly detection in industrial control systems,''
  in \emph{Proc.\ Asia Conference on Computer and Communications Security
  (AsiaCCS 2018)}.\hskip 1em plus 0.5em minus 0.4em\relax {ACM}, 2018, pp.
  525--536.

\bibitem{Narayanan-Bobba18a}
V.~Narayanan and R.~B. Bobba, ``Learning based anomaly detection for industrial
  arm applications,'' in \emph{Proc.\ Workshop on Cyber-Physical Systems
  Security and PrivaCy (CPS-SPC 2018)}.\hskip 1em plus 0.5em minus 0.4em\relax
  {ACM}, 2018, pp. 13--23.

\bibitem{Schneider-Boettinger18a}
P.~Schneider and K.~B{\"{o}}ttinger, ``High-performance unsupervised anomaly
  detection for cyber-physical system networks,'' in \emph{Proc.\ Workshop on
  Cyber-Physical Systems Security and PrivaCy (CPS-SPC 2018)}.\hskip 1em plus
  0.5em minus 0.4em\relax {ACM}, 2018, pp. 1--12.

\bibitem{Ahmed-et_al18a}
C.~M. Ahmed, M.~Ochoa, J.~Zhou, A.~P. Mathur, R.~Qadeer, C.~Murguia, and
  J.~Ruths, ``\emph{NoisePrint}: Attack detection using sensor and process
  noise fingerprint in cyber physical systems,'' in \emph{Proc.\ Asia
  Conference on Computer and Communications Security (AsiaCCS 2018)}.\hskip 1em
  plus 0.5em minus 0.4em\relax {ACM}, 2018, pp. 483--497.

\bibitem{Ahmed-et_al18b}
C.~M. Ahmed, J.~Zhou, and A.~P. Mathur, ``Noise matters: Using sensor and
  process noise fingerprint to detect stealthy cyber attacks and authenticate
  sensors in {CPS},'' in \emph{Proc.\ Annual Computer Security Applications
  Conference (ACSAC 2018)}.\hskip 1em plus 0.5em minus 0.4em\relax {ACM}, 2018,
  pp. 566--581.

\bibitem{Gu-et_al18a}
Q.~Gu, D.~Formby, S.~Ji, H.~Cam, and R.~A. Beyah, ``Fingerprinting for
  cyber-physical system security: Device physics matters too,'' \emph{{IEEE}
  Security {\&} Privacy}, vol.~16, no.~5, pp. 49--59, 2018.

\bibitem{Kneib-Huth18a}
M.~Kneib and C.~Huth, ``Scission: Signal characteristic-based sender
  identification and intrusion detection in automotive networks,'' in
  \emph{Proc.\ {ACM} {SIGSAC} Conference on Computer and Communications
  Security (CCS 2018)}.\hskip 1em plus 0.5em minus 0.4em\relax {ACM}, 2018, pp.
  787--800.

\bibitem{Cardenas-et_al11a}
A.~A. C{\'{a}}rdenas, S.~Amin, Z.~Lin, Y.~Huang, C.~Huang, and S.~Sastry,
  ``Attacks against process control systems: risk assessment, detection, and
  response,'' in \emph{Proc.\ {ACM} Symposium on Information, Computer and
  Communications Security (AsiaCCS 2011)}.\hskip 1em plus 0.5em minus
  0.4em\relax {ACM}, 2011, pp. 355--366.

\bibitem{Adepu-Mathur16a}
S.~Adepu and A.~Mathur, ``Using process invariants to detect cyber attacks on a
  water treatment system,'' in \emph{Proc.\ International Conference on ICT
  Systems Security and Privacy Protection (SEC 2016)}, ser. IFIP AICT, vol.
  471.\hskip 1em plus 0.5em minus 0.4em\relax Springer, 2016, pp. 91--104.

\bibitem{Adepu-Mathur16b}
------, ``Distributed detection of single-stage multipoint cyber attacks in a
  water treatment plant,'' in \emph{Proc.\ {ACM} Asia Conference on Computer
  and Communications Security (AsiaCCS 2016)}.\hskip 1em plus 0.5em minus
  0.4em\relax {ACM}, 2016, pp. 449--460.

\bibitem{Chen-Poskitt-Sun16a}
Y.~Chen, C.~M. Poskitt, and J.~Sun, ``Towards learning and verifying invariants
  of cyber-physical systems by code mutation,'' in \emph{Proc.\ International
  Symposium on Formal Methods (FM 2016)}, ser. LNCS, vol. 9995.\hskip 1em plus
  0.5em minus 0.4em\relax Springer, 2016, pp. 155--163.

\bibitem{Adepu-Mathur18b}
S.~Adepu and A.~Mathur, ``Distributed attack detection in a water treatment
  plant: Method and case study,'' \emph{IEEE Transactions on Dependable and
  Secure Computing}, 2018.

\bibitem{Chen-Poskitt-Sun18a}
Y.~Chen, C.~M. Poskitt, and J.~Sun, ``Learning from mutants: Using code
  mutation to learn and monitor invariants of a cyber-physical system,'' in
  \emph{Proc.\ {IEEE} Symposium on Security and Privacy (S{\&}P 2018)}.\hskip
  1em plus 0.5em minus 0.4em\relax {IEEE} Computer Society, 2018, pp. 648--660.

\bibitem{Choi-et_al18a}
H.~Choi, W.~Lee, Y.~Aafer, F.~Fei, Z.~Tu, X.~Zhang, D.~Xu, and X.~Xinyan,
  ``Detecting attacks against robotic vehicles: {A} control invariant
  approach,'' in \emph{Proc.\ {ACM} {SIGSAC} Conference on Computer and
  Communications Security ({CCS} 2018)}.\hskip 1em plus 0.5em minus 0.4em\relax
  {ACM}, 2018, pp. 801--816.

\bibitem{Giraldo-et_al18a}
J.~Giraldo, D.~I. Urbina, A.~Cardenas, J.~Valente, M.~A. Faisal, J.~Ruths,
  N.~O. Tippenhauer, H.~Sandberg, and R.~Candell, ``A survey of physics-based
  attack detection in cyber-physical systems,'' \emph{{ACM} Computing Surveys},
  vol.~51, no.~4, pp. 76:1--76:36, 2018.

\bibitem{CPS-Datasets}
``{iTrust Labs: Datasets},''
  {\url{https://itrust.sutd.edu.sg/itrust-labs_datasets/}}, 2019, accessed:
  September 2019.

\bibitem{Goh-et_al16a}
J.~Goh, S.~Adepu, K.~N. Junejo, and A.~Mathur, ``A dataset to support research
  in the design of secure water treatment systems,'' in \emph{Proc.\
  International Conference on Critical Information Infrastructures Security
  (CRITIS 2016)}, 2016.

\bibitem{Adepu-Mathur18a}
S.~Adepu and A.~Mathur, ``Assessing the effectiveness of attack detection at a
  hackfest on industrial control systems,'' \emph{IEEE Transactions on
  Sustainable Computing}, 2018.

\bibitem{Goldberg89a}
D.~E. Goldberg, \emph{Genetic Algorithms in Search, Optimization and Machine
  Learning}.\hskip 1em plus 0.5em minus 0.4em\relax Addison-Wesley, 1989.

\bibitem{SWaT-Reference}
``{Secure Water Treatment (SWaT)},''
  {\url{https://itrust.sutd.edu.sg/itrust-labs-home/itrust-labs_swat/}}, 2019,
  accessed: September 2019.

\bibitem{Ahmed-et_al17a}
C.~M. Ahmed, V.~R. Palleti, and A.~P. Mathur, ``{WADI:} a water distribution
  testbed for research in the design of secure cyber physical systems,'' in
  \emph{Proc.\ International Workshop on Cyber-Physical Systems for Smart Water
  Networks (CySWATER@CPSWeek 2017)}.\hskip 1em plus 0.5em minus 0.4em\relax
  {ACM}, 2017, pp. 25--28.

\bibitem{Adepu-Mathur16c}
S.~Adepu and A.~Mathur, ``Water-defense: a method to detect multi-point cyber
  attacks on water treatment systems,'' U.S. provisional application no.
  62/314,6, 2016, provisional patent application no.\ 62/314,6.

\bibitem{Takanen-Demott-Miller18a}
A.~Takanen, J.~D. Demott, and C.~Miller, \emph{Fuzzing for Software Security
  Testing and Quality Assurance}, 2nd~ed.\hskip 1em plus 0.5em minus
  0.4em\relax Artech House, 2018.

\bibitem{Cha-Woo-Brumley15a}
S.~K. Cha, M.~Woo, and D.~Brumley, ``Program-adaptive mutational fuzzing,'' in
  \emph{Proc.\ {IEEE} Symposium on Security and Privacy (S{\&}P 2015)}.\hskip
  1em plus 0.5em minus 0.4em\relax {IEEE} Computer Society, 2015, pp. 725--741.

\bibitem{Zalewski}
M.~Zalewski, ``{American fuzzy lop},'' \url{http://lcamtuf.coredump.cx/afl/},
  2017, accessed: September 2019.

\bibitem{Ghaeini-Tippenhauer16a}
H.~R. Ghaeini and N.~O. Tippenhauer, ``{HAMIDS:} hierarchical monitoring
  intrusion detection system for industrial control systems,'' in \emph{Proc.\
  Workshop on Cyber-Physical Systems Security and Privacy (CPS-SPC
  2016)}.\hskip 1em plus 0.5em minus 0.4em\relax {ACM}, 2016, pp. 103--111.

\bibitem{Goh_et-al17a}
J.~Goh, S.~Adepu, M.~Tan, and Z.~S. Lee, ``Anomaly detection in cyber physical
  systems using recurrent neural networks,'' in \emph{Proc.\ International
  Symposium on High Assurance Systems Engineering (HASE 2017)}.\hskip 1em plus
  0.5em minus 0.4em\relax {IEEE}, 2017, pp. 140--145.

\bibitem{Feng-et_al19a}
C.~Feng, V.~R. Palleti, A.~Mathur, and D.~Chana, ``A systematic framework to
  generate invariants for anomaly detection in industrial control systems,'' in
  \emph{Proc.\ Annual Network and Distributed System Security Symposium (NDSS
  2019)}.\hskip 1em plus 0.5em minus 0.4em\relax The Internet Society, 2019.

\bibitem{Klick-et_al15a}
J.~Klick, S.~Lau, D.~Marzin, J.~Malchow, and V.~Roth, ``Internet-facing plcs as
  a network backdoor,'' in \emph{Proc.\ {IEEE} Conference on Communications and
  Network Security (CNS 2015)}.\hskip 1em plus 0.5em minus 0.4em\relax {IEEE},
  2015, pp. 524--532.

\bibitem{Abbasi-Hashemi16a}
A.~Abbasi and M.~Hashemi, ``Ghost in the {PLC}: Designing an undetectable
  programmable logic controller rootkit via pin control attack,'' in
  \emph{Black Hat Europe}.\hskip 1em plus 0.5em minus 0.4em\relax Black Hat,
  2016, pp. 1--35.

\bibitem{Garcia-et_al17a}
L.~Garcia, F.~Brasser, M.~H. Cintuglu, A.~Sadeghi, O.~A. Mohammed, and S.~A.
  Zonouz, ``Hey, my malware knows physics! {Attacking} {PLCs} with physical
  model aware rootkit,'' in \emph{Proc.\ Annual Network and Distributed System
  Security Symposium (NDSS 2017)}.\hskip 1em plus 0.5em minus 0.4em\relax The
  Internet Society, 2017.

\bibitem{Gers-Schmidhuber-Cummins00a}
F.~A. Gers, J.~Schmidhuber, and F.~A. Cummins, ``Learning to forget: Continual
  prediction with {LSTM},'' \emph{Neural Computation}, vol.~12, no.~10, pp.
  2451--2471, 2000.

\bibitem{Drucker-et_al96a}
H.~Drucker, C.~J.~C. Burges, L.~Kaufman, A.~J. Smola, and V.~Vapnik, ``Support
  vector regression machines,'' in \emph{Proc.\ Advances in Neural Information
  Processing Systems (NIPS 1996)}.\hskip 1em plus 0.5em minus 0.4em\relax {MIT}
  Press, 1996, pp. 155--161.

\bibitem{Chollet-et_al15a}
F.~Chollet \emph{et~al.}, ``Keras,'' \url{https://keras.io}, 2015.

\bibitem{Buitinck-et_al13a}
L.~Buitinck, G.~Louppe, M.~Blondel, F.~Pedregosa, A.~Mueller, O.~Grisel,
  V.~Niculae, P.~Prettenhofer, A.~Gramfort, J.~Grobler, R.~Layton,
  J.~VanderPlas, A.~Joly, B.~Holt, and G.~Varoquaux, ``{API} design for machine
  learning software: experiences from the scikit-learn project,'' in
  \emph{Proc.\ ECML PKDD Workshop: Languages for Data Mining and Machine
  Learning}, 2013, pp. 108--122.

\bibitem{Ruscito}
A.~Ruscito, ``pycomm,'' \url{https://github.com/ruscito/pycomm}, 2019,
  accessed: September 2019.

\bibitem{Supplementary-Material}
``Supplementary material,'' \url{http://sav.sutd.edu.sg/?page_id=3666}, 2019.

\bibitem{Liu-et_al11a}
Y.~Liu, P.~Ning, and M.~K. Reiter, ``False data injection attacks against state
  estimation in electric power grids,'' \emph{{ACM} Transactions on Information
  and System Security}, vol.~14, no.~1, pp. 13:1--13:33, 2011.

\bibitem{Huang-et_al18a}
Z.~Huang, S.~Etigowni, L.~Garcia, S.~Mitra, and S.~A. Zonouz, ``Algorithmic
  attack synthesis using hybrid dynamics of power grid critical
  infrastructures,'' in \emph{Proc.\ {IEEE/IFIP} International Conference on
  Dependable Systems and Networks (DSN 2018)}.\hskip 1em plus 0.5em minus
  0.4em\relax {IEEE} Computer Society, 2018, pp. 151--162.

\bibitem{Urbina-et_al16a}
D.~I. Urbina, J.~A. Giraldo, A.~A. C{\'{a}}rdenas, N.~O. Tippenhauer,
  J.~Valente, M.~A. Faisal, J.~Ruths, R.~Candell, and H.~Sandberg, ``Limiting
  the impact of stealthy attacks on industrial control systems,'' in
  \emph{Proc.\ {ACM} {SIGSAC} Conference on Computer and Communications
  Security (CCS 2016)}.\hskip 1em plus 0.5em minus 0.4em\relax {ACM}, 2016, pp.
  1092--1105.

\bibitem{Sugumar-Mathur17a}
G.~Sugumar and A.~Mathur, ``Testing the effectiveness of attack detection
  mechanisms in industrial control systems,'' in \emph{Proc.\ {IEEE}
  International Conference on Software Quality, Reliability and Security
  Companion (QRS-C 2017)}.\hskip 1em plus 0.5em minus 0.4em\relax {IEEE}, 2017,
  pp. 138--145.

\bibitem{Cardenas-et_al14a}
A.~A. C{\'{a}}rdenas, R.~Berthier, R.~B. Bobba, J.~H. Huh, J.~G. Jetcheva,
  D.~Grochocki, and W.~H. Sanders, ``A framework for evaluating intrusion
  detection architectures in advanced metering infrastructures,'' \emph{IEEE
  Transactions on Smart Grid}, vol.~5, no.~2, pp. 906--915, 2014.

\bibitem{Chowdhury-Johnson-Csallner17a}
S.~A. Chowdhury, T.~T. Johnson, and C.~Csallner, ``{CyFuzz}: {A} differential
  testing framework for cyber-physical systems development environments,'' in
  \emph{Proc.\ Workshop on Design, Modeling and Evaluation of Cyber Physical
  Systems (CyPhy 2016)}, ser. LNCS, vol. 10107.\hskip 1em plus 0.5em minus
  0.4em\relax Springer, 2017, pp. 46--60.

\bibitem{Vigna-et_al04a}
G.~Vigna, W.~K. Robertson, and D.~Balzarotti, ``Testing network-based intrusion
  detection signatures using mutant exploits,'' in \emph{Proc.\ {ACM}
  Conference on Computer and Communications Security (CCS 2004)}.\hskip 1em
  plus 0.5em minus 0.4em\relax {ACM}, 2004, pp. 21--30.

\bibitem{Kang-et_al16a}
E.~Kang, S.~Adepu, D.~Jackson, and A.~P. Mathur, ``Model-based security
  analysis of a water treatment system,'' in \emph{Proc.\ International
  Workshop on Software Engineering for Smart Cyber-Physical Systems (SEsCPS
  2016)}.\hskip 1em plus 0.5em minus 0.4em\relax {ACM}, 2016, pp. 22--28.

\bibitem{Castellanos-Ochoa-Zhou18a}
J.~H. Castellanos, M.~Ochoa, and J.~Zhou, ``Finding dependencies between
  cyber-physical domains for security testing of industrial control systems,''
  in \emph{Proc.\ Annual Computer Security Applications Conference (ACSAC
  2018)}.\hskip 1em plus 0.5em minus 0.4em\relax {ACM}, 2018, pp. 582--594.

\bibitem{McLaughlin-et_al14a}
S.~E. McLaughlin, S.~A. Zonouz, D.~J. Pohly, and P.~D. McDaniel, ``A trusted
  safety verifier for process controller code,'' in \emph{Proc.\ Annual Network
  and Distributed System Security Symposium (NDSS 2014)}.\hskip 1em plus 0.5em
  minus 0.4em\relax The Internet Society, 2014.

\bibitem{Etigowni-et_al16a}
S.~Etigowni, D.~J. Tian, G.~Hernandez, S.~A. Zonouz, and K.~R.~B. Butler,
  ``{CPAC:} securing critical infrastructure with cyber-physical access
  control,'' in \emph{Proc.\ Annual Conference on Computer Security
  Applications (ACSAC 2016)}.\hskip 1em plus 0.5em minus 0.4em\relax {ACM},
  2016, pp. 139--152.

\bibitem{Frehse-et_al11a}
G.~Frehse, C.~L. Guernic, A.~Donz{\'{e}}, S.~Cotton, R.~Ray, O.~Lebeltel,
  R.~Ripado, A.~Girard, T.~Dang, and O.~Maler, ``{SpaceEx}: Scalable
  verification of hybrid systems,'' in \emph{Proc.\ International Conference on
  Computer Aided Verification (CAV 2011)}, ser. LNCS, vol. 6806.\hskip 1em plus
  0.5em minus 0.4em\relax Springer, 2011, pp. 379--395.

\bibitem{Wang-et_al18a}
J.~Wang, J.~Sun, Y.~Jia, S.~Qin, and Z.~Xu, ``Towards 'verifying' a water
  treatment system,'' in \emph{Proc.\ International Symposium on Formal Methods
  (FM 2018)}, ser. LNCS, vol. 10951.\hskip 1em plus 0.5em minus 0.4em\relax
  Springer, 2018, pp. 73--92.

\bibitem{Gao-Kong-Clarke13a}
S.~Gao, S.~Kong, and E.~M. Clarke, ``{dReal}: An {SMT} solver for nonlinear
  theories over the reals,'' in \emph{Proc.\ International Conference on
  Automated Deduction (CADE 2013)}, ser. LNCS, vol. 7898.\hskip 1em plus 0.5em
  minus 0.4em\relax Springer, 2013, pp. 208--214.

\bibitem{Hasuo-Suenaga12a}
I.~Hasuo and K.~Suenaga, ``Exercises in nonstandard static analysis of hybrid
  systems,'' in \emph{Proc.\ International Conference on Computer Aided
  Verification (CAV 2012)}, ser. LNCS, vol. 7358.\hskip 1em plus 0.5em minus
  0.4em\relax Springer, 2012, pp. 462--478.

\bibitem{Lanotte-et_al17a}
R.~Lanotte, M.~Merro, R.~Muradore, and L.~Vigan{\`{o}}, ``A formal approach to
  cyber-physical attacks,'' in \emph{Proc.\ {IEEE} Computer Security
  Foundations Symposium (CSF 2017)}.\hskip 1em plus 0.5em minus 0.4em\relax
  {IEEE} Computer Society, 2017, pp. 436--450.

\bibitem{Kong-et_al16a}
P.~Kong, Y.~Li, X.~Chen, J.~Sun, M.~Sun, and J.~Wang, ``Towards concolic
  testing for hybrid systems,'' in \emph{Proc.\ International Symposium on
  Formal Methods (FM 2016)}, ser. LNCS, vol. 9995.\hskip 1em plus 0.5em minus
  0.4em\relax Springer, 2016, pp. 460--478.

\bibitem{Quesel-et_al16a}
J.~Quesel, S.~Mitsch, S.~M. Loos, N.~Arechiga, and A.~Platzer, ``How to model
  and prove hybrid systems with {KeYmaera}: a tutorial on safety,''
  \emph{International Journal on Software Tools for Technology Transfer},
  vol.~18, no.~1, pp. 67--91, 2016.

\end{thebibliography}

\end{document}